\documentclass[letterpaper, 10 pt,  conference]{ieeeconf}

\usepackage[utf8]{inputenc}
\usepackage{amsmath}
\usepackage{amsfonts}
\usepackage{amssymb}
\usepackage{graphicx}
\usepackage{float}
\usepackage[dvipsnames]{xcolor}
\usepackage{multirow}
\usepackage{todonotes}
\usepackage{comment}
\usepackage{bbding}
\usepackage{amssymb}
\usepackage{pifont}
\usepackage{booktabs}
\usepackage{bbm}
\usepackage{booktabs}
\usepackage{hyperref}

\usepackage{tabularx}
\newcolumntype{L}[1]{>{\raggedright\arraybackslash}p{#1}} 

\newlength{\texttwfontsize}
\newlength{\textssfontsize}
\newcommand{\CN}[1]{\textcolor{red}{CN: #1}}
\newcommand{\LW}[1]{\textcolor{teal}{LW: #1}}
\newcommand{\TDG}[1]{\textcolor{purple}{TdG: #1}}
\newcommand{\THe}[1]{\textcolor{orange}{THe: #1}}
\newcommand{\EB}[1]{\textcolor{green}{EB: #1}}
\newcommand{\EMo}[1]{\textcolor{blue}{EMo: #1}}
\newcommand{\TODO}[1]{\textcolor{gray}{#1}}

\renewcommand{\CN}[1]{}
\renewcommand{\LW}[1]{}
\renewcommand{\TDG}[1]{}
\renewcommand{\THe}[1]{}
\renewcommand{\EB}[1]{}
\renewcommand{\EMo}[1]{}
\renewcommand{\TODO}[1]{}

\def\chapterautorefname~{Chapter~}
\def\sectionautorefname~{Section~}
\def\subsectionautorefname~{Section~}
\def\subsubsectionautorefname~{Section~}
\def\figureautorefname~{Figure~}
\def\tableautorefname~{Table~}

\IEEEoverridecommandlockouts

\title{Fundamental Considerations around Scenario-Based Testing\\for Automated Driving$^*$\thanks{
		$^*$The research leading to these results is partly funded by the German Federal Ministry for Economic Affairs and Energy within the project ``VVM - Verification \& Validation Methods for Automated Vehicles Level 4 and 5'' and by the German Federal Ministry of Education and Research within the project ``TESTOMAT - The Next Level of Test Automation'' under grant agreement No. 01IS17026H.}}

\author{Christian Neurohr$^{1}$, Lukas Westhofen$^{1}$, Tabea Henning$^{1}$, Thies de Graaff$^{1}$, Eike Möhlmann$^{1}$, Eckard Böde$^{1}$\thanks{$^1$OFFIS e.V., Oldenburg, Germany}}

\begin{document}

\maketitle

\begin{abstract}
The homologation of automated vehicles, being safety-critical complex systems, requires sound evidence for their safe operability. 
Traditionally, verification and validation activities are guided by a combination of ISO 26262 and ISO/PAS 21448, together with distance-based testing. 
Starting at SAE Level 3, such approaches become infeasible, resulting in the need for novel methods. Scenario-based testing is regarded as a possible enabler for verification and validation of automated vehicles. 
Its effectiveness, however, rests on the consistency and substantiality of the arguments used in each step of the process. 
In this work, we sketch a generic framework around scenario-based testing and analyze contemporary approaches to the individual steps. 
For each step, we describe its function, discuss proposed approaches and solutions, and identify the underlying arguments, principles and assumptions. 
As a result, we present a list of fundamental considerations for which evidences need to be gathered in order for scenario-based testing to support the homologation of automated vehicles.
\end{abstract}
	
\section{Introduction}\label{sec:intro}
% !TeX spellcheck = en_US

The introduction of automated vehicles (AVs) to public roads promises many benefits. 
These include a reduced number of accidents caused by driver errors (safety), increased efficiency of the transport system (environment), increased spare time for users (comfort) and mobility for elderly and impaired users (social inclusion) \cite{ertrac2019}.
At higher levels of automation, in particular at SAE Level $\ge 3$, AVs become complex autonomous systems operating in open context, i.e.\ dealing with unstructured real world environments. 
Thus, the validation and approval of AVs pose an enormous challenge \cite{poddey_validation_2019}. 
As a newly emerging and safety-critical technology, rigorously proving safe operability and strictly avoiding accidents is crucial for societal acceptance.

The process of assuring functional safety for conventional road-vehicles is guided by the ISO 26262 \cite{iso26262}. 
This process has been extended by the ISO/PAS 21448 \cite{iso21448} to include the complementary aspect of the safety of the intended functionality (SOTIF), at least for SAE Level $\le 2$.
In the automotive domain, testing is an important part of verification and validation and has traditionally used distance-based statistical arguments. 
However, for AVs, distance-based approaches to testing become infeasible due to the vast amount of distance that needs to be covered \cite{kalra2016driving} \cite{wachenfeld2016release}. 
Instead, other approaches begin to emerge, such as 'Responsibility Sensitive Safety' (RSS), which is a formal model that relies on defining a safety envelop around the vehicle \cite{shalev-shwartz_formal_2018}. 
However, it is questionable whether this formal model can actually be implemented in the real world \cite{koopman2019autonomous}. 
A novel approach, which has been explored by recent research projects such as PEGASUS \cite{PEGASUSmethod} and ENABLE-S3 \cite{ENABLES3report}, is to perform scenario-based testing (SBT). 
In SBT, testing is performed by deriving relevant test cases from a manageable set of scenario classes.

In this paper, we perform a deep dive into the fundamental considerations that must be taken into account in order for SBT to generate a meaningful contribution to the verification and validation of AVs. 
The goal is to identify underlying principles and assumptions which are essential to the successful realization of SBT for automated driving. 
We discuss relevant terminology, notation and a general framework around SBT in \autoref{sec:app}. 
The introduced framework provides the structure for \autoref{sec:assump}. 
For each process step, we first describe the state of the art. Secondly, we semi-formally examine the fundamental arguments, principles, and assumptions that are inherent to the aforementioned contemporary approaches. 
The core results of our analysis are summarized in \autoref{sec:res} as a condensed table, which may serve as an impulse for future research.

\section{Scenario-Based Testing}\label{sec:app}
% !TeX spellcheck = en_US

\begin{figure*}[t]
	\centering
	\includegraphics[width=\textwidth-2em]{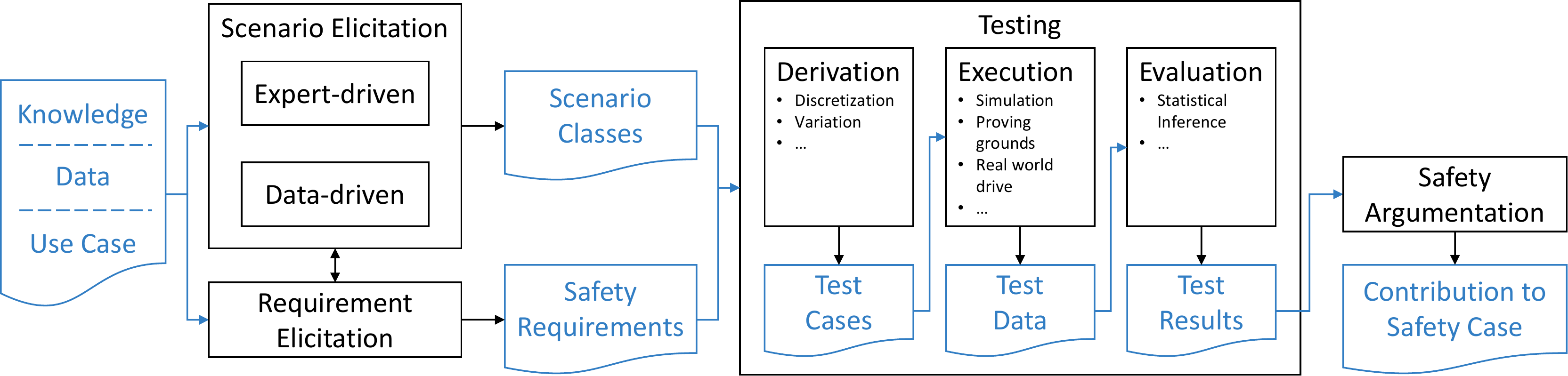}
	\caption{Simplified framework around scenario-based testing based on \cite{PEGASUSmethod} and \cite{ENABLES3report}. Rectangular boxes represent process steps, whereas curved boxes depict artifacts.}\label{fig:sb-testing-workflow}
\end{figure*}

Verification and validation serves the purpose of assuring properties of interest, such as safety and security, with respect to both the specified and intended purpose of the system. 
In the automotive domain, testing is defined as the 'process of planning, preparing, and operating or exercising an item or an element to verify that it satisfies specified requirements, to detect anomalies, and to create confidence in its behavior' \cite{iso26262}. 
In SBT, the central artifact for the operation is the scenario. 
Such testing approaches are pursued as state of the art in the AV testing community \cite{stellet_testing_2015} \cite{tahir_coverage_2020}.

\subsection{Terminology and Notation}
\label{subsec:tandn}
A standardized definition of the term scenario in the context of verification and validation of automated vehicles is presented in ISO/PAS 21448 \cite{iso21448}, which we briefly introduce alongside a semi-formalism that will be used later on.

A \emph{scenario} describes the development of scenes over time, thus building a temporal sequence. 
A \emph{test case} is defined as a scenario enriched with pass/fail criteria suitable for test evaluation \cite{iso21448}. 
Menzel et al. propose the qualifications of functional, logical and concrete scenarios \cite{menzel2018scenarios}. 
At the highest level of abstraction, a \emph{functional scenario} is described semantically using natural language.

Formalizing a functional scenario, a \emph{logical scenario} $L$ can be represented by a state space and its interrelations. 
It includes a list of relevant parameters $\mathcal{A}_L = (\alpha_1, \dots, \alpha_n)$ and their value ranges $\mathcal{R}_L = (V_1, \dots, V_n)$ with $V_i \subseteq \mathbb{R}$ for $1 \leq i \leq n$. 
The state space of $L$ can be described by $\mathcal{V}_L = V_1 \times \dots \times V_n \subseteq \mathbb{R}^n$. Optionally, correlations between parameters and numeric constraints can be added. 
Probability distributions can be attached to the state space or the value ranges. 
We denote the set of all logical scenarios by $\mathcal{L}$.

A \emph{concrete scenario} $C$ requires the assignment of a single value to each parameter. 
It can be obtained from a logical scenario $L$ by instantiating all parameters $\alpha_i \in \mathcal{A}_L$ with some $v_i \in V_i$, i.e. selecting a $v \in \mathcal{V}_L$. 
We denote the set of all concrete scenarios by $\mathcal{C}$. Analogously, $\mathcal{C}_L$ is the set of all concrete scenarios derived from a logical scenario $L$. 
The process of instantiating a set of logical scenarios to concrete scenarios can be seen as a map $\mathcal{I}: \mathcal{L} \to \mathcal{C}, L \mapsto C \in \mathcal{C}_L$ w.r.t. the relations, constraints and distributions of $L$.

\subsection{General Framework}

In order to generate the aforementioned test cases and to interpret the test results, various steps are commonly executed up- and downstream of the actual testing process. 
Thus, it is crucial to embed SBT into a larger framework. 
\autoref{fig:sb-testing-workflow} shows such an abstract framework around SBT, mainly based on the ENABLE-S3 scenario-based verification and validation process \cite{ENABLES3report} and the PEGASUS method \cite{PEGASUSmethod}. 
Note that this framework is not meant to be a core contribution of our work, but rather an organizational tool for the following considerations obtained from incorporating various existing approaches.
%Note that this framework is not a core contribution of our work but rather presents a common denominator of various existing publications, with the goal of enabling an organization of our considerations.
For example, the framework is also consistent with the publications by Huang et al. \cite[Figure 8]{huang_autonomous_2016} and Kalisvaart et al. \cite[Figure 7]{kalisvaart_using_2020}. 

The first step, \emph{scenario elicitation}, consists of deriving adequate scenario classes, e.g. in the form of logical scenarios. 
The \emph{requirement elicitation} step complements these classes by a set of safety requirements that shall be satisfied by the system in the identified scenarios. 
\emph{Testing} then evaluates the system's compliance to the requirements w.r.t. the scenario classes. 
Well-established concepts around testing can be found in standards such as the ISO/IEC/IEEE 29119 \cite{iso29119} and ISO 26262 \cite{iso26262}. 
Within this test process, test cases are initially derived from the scenario classes and safety requirements during \emph{test derivation}. 
Subsequently, \emph{test execution} is performed using an appropriate test bench, e.g. either virtually, physically or in a combination. 
The executed tests are assessed in a \emph{test evaluation} step.
The results of this whole process are then used in a subsequent, overarching \emph{safety argumentation}, which contributes to the safety case.

\section{Fundamental Considerations}\label{sec:assump}
% !TeX spellcheck = en_US

Structured along this framework, we examine arguments, principles and assumptions that, in the authors' opinion, are fundamental to the SBT process. 
We initially present some general considerations applicable to the overall process in \autoref{sec:generalconsiderations}. 
Then, in \autoref{sec:scenarioelicitation} to \autoref{sec:testevaluation}, we conduct a two-part examination for each process step: we first give an overview of the state of the art which is followed by a structured analysis of the underlying fundamental considerations. 
The key insights are marked with an ID, linking them to \autoref{tab:considerations}.
We omit an in-depth analysis of the safety argumentation, which will be subject to further research.

\subsection{General Considerations}
\label{sec:generalconsiderations}

SBT requires a unified understanding of what the term scenario means. As mentioned in \autoref{subsec:tandn}, ongoing standardization activities work towards this unified understanding, but there remain open questions on the exact definition and usage of scenarios and what its qualifications describe.
For example, the ISO/PAS 21448 uses logical scenarios in the test phase while Menzel et al. propose to use concrete scenarios \cite{menzel2018scenarios}. 
Depending on these definitions, the technical realizations of these qualifications, e.g. OpenDRIVE\footnote{\url{asam.net/standards/detail/opendrive}} and OpenSCENARIO\footnote{\url{asam.net/standards/detail/openscenario}}, need to be sufficiently expressive and unambiguous such that scenario descriptions lead to test cases without loss or misinterpretation of information (\texttt{G1}).

Second, SBT often assumes smoothness of certain properties of interest, e.g. of criticality metrics, behaviors or trajectories, under variation of the parameters, i.e.\ slight variations of the parameters lead to slight changes of the considered property (\texttt{G2}). 
This smoothness can be exploited, e.g. to extrapolate knowledge of a single scenario to a set of scenarios or to explore the close proximity -- in the sense of parameters -- of a scenario in order to estimate the direction of more challenging scenarios.
In both cases, this increases the certainty of the inference from finitely many test results to sets of scenarios when analyzing complex systems.
However, Poddey et al.~\cite{poddey_validation_2019} argue that AVs are an instance of complex systems which rely heavily on sensor technologies for the perception of their open context. This poses difficulties, as
processing the sensory input often uses machine learning. These algorithms are prone to misclassifications as a result of minor input changes \cite{akhtar2018threat}, therefore smoothness of such classification algorithms cannot easily be assumed. 

Since the open context in which AVs are supposed to operate is constantly evolving, the framework around SBT has to be adaptable.
A complete set of scenario classes and testing methods at the time of elicitation may be incomplete at time of deployment due to a number of unforeseeable reasons, so called \emph{unknown unknowns} \cite{iso21448} \cite{lakkaraju2017identifying}. 
These range from novel objects or vehicles appearing in traffic over newly introduced traffic regulations up to massive structural changes in traffic due to human drivers reacting to the introduction of AVs. 
Hence, an integrated update process and constant monitoring of AVs during their life cycle are inevitable for SBT (\texttt{G3}).

\subsection{Scenario Elicitation}
\label{sec:scenarioelicitation}

% DESCRIPTION
Most published frameworks follow an expert- or data-driven approach to elicit functional or logical scenarios.
% EXPERT-DRIVEN
Expert-driven approaches use a manifestation and formalization of expert knowledge about the real world. 
Proposed methods include the direct use of laws, regulations, system specifications, as well as already existing logical scenarios \cite{ponn_optimization-based_2019}. 
As an a priori formalization step, it has been proposed to incorporate such knowledge into ontologies \cite{bagschik2018ontology}. 
Based on such formalized knowledge, experts can conduct safety analyses to identify hazardous scenarios \cite{neurohr2020}.
%DATA-DRIVEN
Scenarios can also be elicited in a data-driven manner. 
For example, large-scale observational studies can serve as a means to identify hazardous events, e.g. by learning a neural network \cite{gruner_spatiotemporal_2017}. 
Those can in turn be transformed into logical scenarios, e.g. by using the challenger framework \cite{weber2019framework}, and can additionally be attached with probability distributions from the observed data \cite{putz_system_2017}.
Such a process can be enhanced by extending the already existing data basis \cite{hauer_did_2019}. 
Besides investigating large data sets, one can also closely investigate a concrete real-world drive \cite{zofka_data-driven_2015}, which then allows for a parametrization into a logical scenario.
% EXPERT- & DATA-DRIVEN
Finally, there exist attempts to combine expert- and data-driven frameworks. 
In top-down approaches, logical scenarios are classified by experts, e.g. on a keyword-based scheme \cite{menzel_functional_2019}, and the attached probability distributions are derived from real-world data. 
Orthogonally, bottom-up approaches use expert analysis on data bases, e.g. a Successive Odds Ratio Analysis \cite{watanabe_scenario_2019} or a classification tree approach \cite{bach_test_2017}, to classify observed concrete scenarios.

% ANALYSIS
% EXPERT-DRIVEN
Expert-driven scenario elicitation necessitates defining a set of relevant real-world phenomena to be considered implicitly or explicitly, e.g. via an ontological representation.
This approach is prone to creating a gap between identified and relevant phenomena. While this gap can be controlled by using systematic methods and made visible by explicit representations, completeness guarantees cannot be given. Moreover, experts lack the combinatorial power for the exhaustive exploration of the open context, increasing the risk of an incomplete set of scenario classes. In that regard, automated analyses are able to complement an expert-based approach (\texttt{SE1}).
% DATA-DRIVEN
Most data-driven approaches rely on the existence of a sufficiently large data basis. 
It is assumed that relevant phenomena leave traces in recorded data, for example as causes of accidents in observational studies. 
In this case, relevant phenomena have to be present either directly in the data model or need to be manually identified by examination of the recorded scenarios (\texttt{SE2}).
If the data basis is extended by collecting more data sets, the choice of observation is up to expert judgment and arguments must be made as to why this choice leads to scenario classes that enable SBT. 
Regarding the location of data collection, experts must argue that the chosen locations are in some form representative for the targeted operational design domain. 
This applies particularly to the validity of measured real-world distributions and their generalizability. 
For collecting data and evaluating concrete real-world drives, it has to be decided which entities and relations shall be observed in the real world. 
Additionally, appropriate tools and measurement sensors with sufficient accuracy have to be chosen purposefully and explicitly, in order for a downstream safety argumentation to reason over the validity of such decisions (\texttt{SE3}).
% GENERAL
Both expert- and data-driven approaches have in common that they either elicit an arguably complete classification of the scenario space, or need to reason on why the exclusion of certain scenarios is applicable in the current context (\texttt{SE4}).
While a complete classification of scenarios that covers the entire scenario space is essential for SBT, such a decomposition does not reduce the overall test effort. 
However, if sound arguments can be made why some classes can be excluded altogether, omitting these classes significantly reduces the required test effort later on. 

\subsection{Requirement Elicitation}

%DESCRIPTION
Of particular importance for any safety-relevant application is a systematic safety process that includes identification of hazards and risks, derivation of safety goals and decomposition of the high-level safety goals into safety requirements on item and component level \cite{klamann_defining_2019} \cite{winner2019requirements}.
For systems up to SAE Level 2, this is well covered by ISO 26262 and ISO/PAS 21448. 
For  SAE Level $\ge 3$, scenarios appear to be an ideal starting point for the derivation and elicitation of acceptance criteria. 
Junietz et al. examine how such a quantified requirement definition can take place based on acceptable risk levels \cite{junietz_macroscopic_2019}. 

%ANALYSIS
The process of identifying and specifying requirements is inherently bound to the problem of \textit{correctness} (does the specification match the actual requirement), \textit{completeness} (are all safety goals fully covered), \textit{consistency} (the absence of contradictions) and \textit{validity} (are we specifying a useful requirement in the first place)
(\texttt{RE1}).
Furthermore, for requirements tested virtually, specific arguments must be made with regards to measurability and computability. 
Here, the number of virtually executed scenarios can grow extremely large. 
Thus, the key elements of requirements need to be observable and measurable in the virtual environment
(\texttt{RE2}).

\subsection{Test Derivation}\label{subsec:testder}

\subsubsection{Discretization}
During test derivation, a discretization step for a class $L$ to a discrete class $\tilde{L}$ is usually performed to reduce the test effort to a finite number of concretizations. 
This step can be seen as a map $\mathcal{C}_L \mapsto \mathcal{C}_{\tilde{L}}$ with $|\mathcal{C}_{\tilde{L}}| < \infty$. 
If the state space $\mathcal{V}_L$ of a logical scenario $L$, i.e. a whole scenario class, is fully described using a set of bounded value ranges $\mathcal{V}_L = V_1 \times \dots \times V_n \subset \mathbb{R}^n$, some of these parameters describe measurable continuous real-world quantities. Examples contain the speed of an object and rainfall quantity. 
Mapping each continuous value range to a finite, discrete one, denoted by $V_i \mapsto \tilde{V}_i$, results in the aforementioned logical scenario $\tilde{L}$ with a finite, discrete state space $\mathcal{V}_{\tilde{L}} = \tilde{V}_1 \times \dots \times \tilde{V}_n \subsetneq \mathcal{V}_L$.

While overly fine discretizations may produce many redundant test cases \cite{ponn_optimization-based_2019}, any discretization step leads to gaps in the original state space, possibly resulting in undiscovered safety violations \cite{amersbach2019functional}. 
One approach defines the map $\mathcal{C}_L \mapsto \mathcal{C}_{\tilde{L}}$ according to probability distributions of the parameters \cite{eberle2020simulation}, obtained from either synthesized or real-world data. 
In both cases, however, the question of the validity of the data arises. 
Discretization according to distributions is more sophisticated than equidistant splitting and leads to scenarios that resemble reality more closely, but its usefulness depends on the fundamental assumption that 'probability of occurrence of a scenario correlates with its relevance in terms of the safety argumentation' \cite{eberle2020simulation}. 
Whether this assumption is justified depends on the paradigm used for the safety argumentation \cite{brade2020}. 
We note that unlikely parameter combinations could lead to critical outcomes. 
Thus, discretization might lead to dismissing rare, critical scenarios due to preeliminated parameter combinations (\texttt{TD1}).

\subsubsection{Variation Methods}

In order to generate test cases for a given logical scenario $L$, it is necessary to instantiate the parameters with concrete values~\cite{menzel2018scenarios}.
For efficient testing, a systematic approach for instantiation is required. This can be done either deterministically, e.g.\ by stepping (equidistantly) through $\mathcal{V}_L$ or performing Boundary Value Analysis~\cite{sippl_simulation_2016}, or stochastically, e.g.\ using Monte Carlo methods \cite{akagi_risk-index_2019}. 
If we attach probability distributions to each parameter value range $V_i$ (or a joint distribution defined on $\mathcal{V}_L$), the instantiation $\mathcal{I}$ can be seen as a random variable that assigns concrete values $v_i \in V_i$ to each parameter $\alpha_i \in \mathcal{A}_L$ through sampling.
Then, one may sample $\mathcal{I}$ according to the real-world distributions of the parameters w.r.t. all constraints and interrelations of $L$. 
It is likely that these distributions are approximated from either locally measured naturalistic data  \cite{wagner_using_2018} or from valid simulated data \cite{akagi_risk-index_2019} \cite{eberle2020simulation}. 
Both cases assume that sampling $\mathcal{I}$ according to real-world distributions enhances the value of test cases for the safety argument. 
 
A different approach argues that scenarios being critical w.r.t. a suitable metric offer the most value for the safety argument and should therefore be tested preferably \cite{junietz2018criticality}. 
However, as these rarely occur naturally \cite{junietz_macroscopic_2019}, one has to increase their significance during the sampling process. 
This can be realized by sampling $\mathcal{I}$. Here, a criticality metric $\kappa\colon \mathcal{S}(\mathcal{C}_L) \to [0,1]$, that maps a discrete time series of values $T$ obtained from a simulation engine $\mathcal{S}$ to a scalar measuring the criticality of a concrete scenario $C \in \mathcal{C}_L$, can be optimized.  
This criticality-guided approach shifts much of the methodical burden to the metrics. First, a criticality metric $\kappa$ has to reflect the real-world criticality accurately \cite{junietz2018criticality}  (\texttt{TD2}).
A single metric is unlikely to capture all types of critical phenomena so that several metrics in conjunction, each capturing different aspects of criticality, are required \cite{hallerbach_simulation-based_2018}. 
Moreover, each metric $\kappa$ is evaluated on a time series $T = \mathcal{S}(C)$ depending on $\mathcal{S}$, where $C = \mathcal{I}(L)$ is described by the parameters $\mathcal{A}_L$.  
Thus, sampling $\mathcal{I}$ in order to optimize $\kappa$ can only capture real-world criticalities that are (i) actively influenced by the parameters $\mathcal{A}_L$ and (ii) accurately computed by $\mathcal{S}$. 
Clearly, (i) depends on the scenario description language and (ii) hinges on the validity of the simulation environment and the utilized models. 
Influencing factors for real-world criticality whose effects are not conveyed by this process can thus not be revealed.
Simple metrics such as TTC incorporate only few parameters, but can be optimized efficiently. More involved metrics cover more influencing factors at the cost of higher computational effort \cite{junietz_criticality_2018}. 
When defining a new metric, e.g.\ by combining known metrics and taking a weighted sum, the complexity of the associated optimization problem is determined by its mathematical properties, e.g. being continuous, differentiable, or convex. 

For each criticality metric $\kappa$ a threshold $c_{\kappa} \in [0,1]$ is required in order to label a time series $T=\mathcal{S}(C)$ as critical whenever $\kappa(T) \ge c_{\kappa}$  (\texttt{TD3}).
Junietz et al. \cite{junietz2017metrik} suggest fitting thresholds based on manually annotated scenarios which are then used to learn binary classifiers. 
Criticality thresholds can be used to define the \emph{critical subspace} $\mathcal{C}_{L, \kappa,\mathcal{S}} = \lbrace C \in \mathcal{C}_L \mid \kappa(\mathcal{S}(C)) \ge c_{\kappa} \rbrace \subset \mathcal{C}_L$ w.r.t. $\kappa$ and $\mathcal{S}$. The question arises whether optimization algorithms reliably cover the set
$\mathcal{C}_{L, \kappa,\mathcal{S}}$ by finding critical concrete scenarios $C \in \mathcal{C}_{L, \kappa,\mathcal{S}}$ from all of its connected components.
If components of $\mathcal{C}_{L, \kappa,\mathcal{S}}$ are missed entirely, one can no longer argue about testing all critical instantiations of $L$  (\texttt{TD4}).
In order to prevent this, optimization algorithms may have to be executed repeatedly, using varying step lengths or starting points. 
In any case, the dependency on the simulation engine $\mathcal{S}$ remains.

\subsection{Test Execution}

% !TeX spellcheck = en_US

\begin{table*}[t]
	\renewcommand{\arraystretch}{1.4}
	\caption{Overview of the examined fundamental considerations around scenario-based testing. Examples of relevant literature are stated for each process step.}
	\label{tab:considerations}
	\begin{center}
		\begin{tabular}{lp{10.5cm}ll}
\toprule
\textbf{ID}	& \textbf{Consideration} & \textbf{Process Step} & \textbf{Literature}\\
\midrule
\texttt{G1}		& The term scenario, its qualifications and technical realizations are sufficiently expressive and unambiguous.		& General					& \multirow[t]{3}{1.8cm}{{\cite{poddey_validation_2019} \cite{iso21448} \cite{menzel2018scenarios} \cite{akhtar2018threat}}} \\
\texttt{G2}		& If smoothness of a property of interest of the system is assumed, this assumption is asserted.				&&\\ 
\texttt{G3}		& The SBT process is adaptable to real-world changes.																&&\\[0.1cm]\cmidrule(lr){1-4}
\texttt{SE1}	& An expert-based approach is executed systematically and supported by automation. 									& Scenario Elicitation		& \multirow[t]{4}{1.8cm}{{\cite{ponn_optimization-based_2019} \cite{bagschik2018ontology} \cite{neurohr2020} \\ \cite{gruner_spatiotemporal_2017} \cite{weber2019framework} \cite{putz_system_2017} \\ \cite{hauer_did_2019} \cite{zofka_data-driven_2015} \cite{menzel_functional_2019} \\ \cite{watanabe_scenario_2019} \cite{bach_test_2017}}} \\
\texttt{SE2}	& An identification of all relevant phenomena is facilitated. 														&&\\
\texttt{SE3}	& A data-driven approach uses representative measurement locations, devices and valid probability distributions.	&&\\
\texttt{SE4}	& A decomposition of the test space into scenario classes is complemented either by evidence for its completeness or an argumentation for the omission of classes.&&\\[0.1cm]\cmidrule(lr){1-4}
\texttt{RE1}	& Identified requirements are correct, consistent, complete and valid.												& Requirement Elicitation 	& \multirow[t]{2}{1.8cm}{\cite{klamann_defining_2019}  \cite{winner2019requirements} \cite{junietz_macroscopic_2019}} \\
\texttt{RE2}	& Key elements of requirements are observable and measurable in the environment.									&&\\[0.1cm]\cmidrule(lr){1-4}
\texttt{TD1}	& Discretization does not preeliminate valuable test cases.															& Test Derivation			& \multirow[t]{4}{1.8cm}{{\cite{menzel2018scenarios} \cite{ponn_optimization-based_2019} \cite{junietz_macroscopic_2019} \cite{amersbach2019functional} \cite{eberle2020simulation} \cite{brade2020}  \cite{sippl_simulation_2016} \cite{akagi_risk-index_2019} \cite{wagner_using_2018} \cite{junietz2018criticality} \cite{hallerbach_simulation-based_2018} \cite{junietz_criticality_2018} \cite{junietz2017metrik}}} \\
\texttt{TD2}	& Utilized criticality measures are validated.																		&&\\
\texttt{TD3}	& Suitable thresholds for criticality measures are employed.														&&\\
\texttt{TD4}	& The critical subspace is explored systematically.																	&&\\[0.1cm]\cmidrule(lr){1-4}
\texttt{TX1}	& The simulation environment, all models and their interactions are validated.										& Test Execution			& \multirow[t]{1}{1.8cm}{{\cite{huang_autonomous_2016} \cite{hakuli2015virtuelle} \cite{steimle_method_2019} \cite{sargent2010verification} \cite{bode2018efficient}}} \\
				&																													&&\\[0.1cm]\cmidrule(lr){1-4}
\texttt{TE1}	& Test results are aggregated into a statement that supports the safety case.										& Test Evaluation			& \multirow[t]{2}{1.8cm}{{\cite{brade2020} \cite{groh_towards_2017} \cite{pegasusposter28} \cite{gerwinn2019statistical}}} \\
\texttt{TE2} 	& Statements about test coverage of scenario classes are derived using sound statistical arguments.					&&\\
\bottomrule
		\end{tabular}
	\end{center}
\end{table*}

Depending on the stage of the development process, there exists a diverse set of test execution methods \cite{huang_autonomous_2016}. 
An essential strategy is mixed virtual-physical testing, i.e.\ replacing physical components with virtual ones \cite{hakuli2015virtuelle}. 
For each involved component and simulation model, a test method needs to be chosen that provides valid test results \cite{steimle_method_2019}.

In order for virtual testing to replace physical testing, the corresponding virtual models have to be verified and validated against their physical counterparts \cite{sargent2010verification}. 
The validity of a simulation environment depends on the validity of many individual models (e.g., environment models, behavior models, sensor models) as well as their interactions.
Gathering relevant real-world data for virtual models is a key challenge in such a validation process, which also depends on the employed notion of validity (cf. \cite{bode2018efficient}). 
Thus, the question how to validate simulation environments remains subject to current research (\texttt{TX1}).

\subsection{Test Evaluation}
\label{sec:testevaluation}

After executing test cases, they need to be evaluated according to safety requirements. 
For SBT, these requirements can appear in the form of thresholds for criticality metrics, being evaluated using a finite sequence of measurements during the evolution of a concrete scenario \cite{groh_towards_2017}. 
It is possible to judge compliance of real-world tests with qualitative requirements using experts. 
For virtual batch-testing of scenarios, however, qualitative requirements need to be mapped to criticality metrics and thresholds.
PEGASUS suggested a 4-stage-process for test case evaluation that evaluates the aspects safety distances, absence of collision, causality and mitigation \cite{pegasusposter28}. 
Instead of binary test results, \cite{brade2020} proposes test case evaluation on a more detailed ordinal scale, incorporating knowledge about possible discrepancies between the anticipated and actual category of a test case depending on the underlying paradigm.

Formalizing test evaluation, we assume that executing a concrete scenario $C$ instantiated by $\mathcal{I}$ from $L$ results in a discrete time series of values $T$. This time series can either be obtained by a simulation engine $\mathcal{S}$ or by real-world measurements. We focus on the former here, i.e.\ $C = \mathcal{I}(L)$ and $T = \mathcal{S}(C)$. A pass/fail-criterion $\kappa(\cdot) < c_{\kappa}$ is denoted as $\mathcal{R}$ w.r.t. a suitable criticality metric $\kappa$ and threshold $c_{\kappa}$ for $L$. 
The results of testing $N$ instantiations $C_1,\dots,C_N$ of $L$, obtained from $\mathcal{R}$, need to be aggregated. A key challenge is to obtain a useful aggregate statement to support the safety case (\texttt{TE1}). While a simple success ratio $\frac{1}{N} \sum_{i=1}^N \mathbf{1}_{\{S(C_i) \models \mathcal{R}\}}$ can be adequate for comfort functions, it might not be a good choice for testing safety goals. 
This challenge becomes even more complex when evaluating multiple metrics.

The map $\mathcal{I}$ and also $\mathcal{S}$ -- for a probabilistic simulation -- can be seen as stochastic processes. Their composition with a  criticality metric $\kappa$ is a random variable $X =  \kappa \circ \mathcal{S} \circ \mathcal{I}\colon \mathcal{L} \to [0,1]$.
For test evaluation of an entire scenario class $L$, for which not all instances can be tested, we need to make a statistical statement about the system under test failing a test coming from $L$ (\texttt{TE2}).
A possible strategy is to estimate the probability of a random scenario derived from $L$ being critical w.r.t. $\kappa$, i.e. $p=p(\kappa,L,\mathcal{S},\mathcal{I}) = P(X \ge c_{\kappa})$. 
The true distribution of $p$ is unknown which requires an estimate $\hat{p}$. 
Assuming the tested system is well-engineered and the threshold $c_{\kappa}$ is well-chosen, we can expect $p$ to be very small. 
As explained in \autoref{subsec:testder}, advanced Monte Carlo methods can be applied to obtain an estimate $\hat{p}$. 
A confidence statement about $p$ using $\hat{p}$ can then be interpreted as a statistical statement about test coverage based on the generated samples \cite{gerwinn2019statistical}.

\section{Results}\label{sec:res}
% !TeX spellcheck = en_US

As the result of our analysis, we present \autoref{tab:considerations}, which
summarizes the fundamental considerations from \autoref{sec:assump}, marked with an identifier. These considerations are grouped according to the steps of the SBT process, as depicted by \autoref{fig:sb-testing-workflow}, and each step is annotated with examples of the relevant literature.

In order to obtain a valid safety case for the release of an AV, a coherent well-structured safety argumentation is required. Such a safety case needs to argue why and how the provided evidence satisfies the high-level safety requirements. One integral part of this evidence are the results of the testing process. In this regard, \autoref{tab:considerations} provides a collection of considerations for which meaningful evidences need to be gathered in order for SBT to support a safety case.
However, this collection is not exhaustive and additional considerations may be necessary to establish a valid safety case.

\section{Conclusion}\label{sec:conc}
% !TeX spellcheck = en_US

A comprehensive review and analysis of the literature concerning SBT for automated vehicles was performed. We presented numerous arguments, principles and assumptions that are fundamental to the automotive SBT approach. For each step of the process, we analyzed the strengths and weaknesses of the most promising contemporary approaches in order to uncover potential gaps and inconsistencies. As a result, we obtained a collection of fundamental considerations that need to be substantiated with evidences.
% and are possible candidates for further research objectives.
It is subject to further research to provide approaches that generate these evidences reliably. Finally, an exploration of additional fundamental considerations complementing our collection is likely to be necessary.

\bibliographystyle{ieeetr}
\bibliography{Literature,Zotero}
\end{document}